\documentclass[pre,reprint,superscriptaddress,showpacs,longbibliography]{revtex4-1}
\usepackage{graphicx}
\usepackage{amsmath} 
\usepackage{txfonts} 
\usepackage{color}

\def\lett#1{(\textbf{#1})}

\newcommand{\pdfrac}[2]{\ensuremath{{\partial #1}/{\partial #2}}}
\def\Pr#1{P\left(#1\right)}
\def\PrPA#1{P_\mathrm{PA}\left(#1\right)}
\def\PrCAn{P_\mathrm{CA}}
\def\PrCA#1{\PrCAn\left(#1\right)}
\def\cbar{\ensuremath{\bar{c}}}
\def\kbar{\ensuremath{\bar{k}}}

\begin{document}

\title{Natural Emergence of Clusters and Bursts in Network Evolution}

\author{James P.~Bagrow}
\email{bagrowjp@gmail.com}
\author{Dirk Brockmann}
\affiliation{Engineering Sciences and Applied Mathematics, Northwestern University, Evanston, Illinois 60208, USA}
\affiliation{Northwestern Institute on Complex Systems, Northwestern University, Evanston, Illinois 60208, USA}

\date{June 24, 2013}

\begin{abstract} 
    Network models with preferential attachment, where new nodes are injected
    into the network and form links with existing nodes proportional to their
    current connectivity, have been well studied for some time. Extensions have
    been introduced where nodes attach proportionally to arbitrary fitness
    functions.  However, in these models, attaching to a node always increases
    the ability of that node to gain more links in the future. We study network
    growth where nodes attach proportionally to the clustering coefficients, or
    local densities of triangles, of existing nodes. Attaching to a node
    typically lowers its clustering coefficient, in contrast to preferential
    attachment or rich-get-richer models. This simple modification naturally
    leads to a variety of rich phenomena, including aging, non-Poissonian bursty
    dynamics, and community formation.  This theoretical model shows that
    complex network structure can be generated without artificially imposing
    multiple dynamical mechanisms and may reveal potentially overlooked
    mechanisms present in complex systems. 
\end{abstract}
\pacs{87.23.Ge, 87.10.-e, 89.75.Hc, 89.75.Fb, 89.20.Hh}

\maketitle

Growing network models have been introduced to study the topological evolution
of systems such as citations between scientific
articles~\cite{price1976general,de1986little,redner2005citation,Medo:2011eu},
protein interactions in various
organisms~\cite{eisenberg2003preferential,albert2005scale}, the World Wide
Web~\cite{Barabasi:1999uu}, and more~\cite{PhysRevLett.85.5234,Newman:2010ur}.
Meanwhile, recent interest has been drawn towards understanding not simply the
topology of these systems or how the individual system elements interact, but also
the temporal nature of these interactions~\cite{Holme2012}.  For example,
studies of the burstiness of human
dynamics~\cite{barabasi2005origin,goh2008burstiness}, whether by letter
writing~\cite{oliveira2005human} or mobile phone
usage~\cite{1367-2630-14-1-013055}, have advanced our knowledge of how information
spreads~\cite{granovetter1973strength,granovetter1978threshold,watts2002simple}
through systems mediated by such
dynamics~\cite{PhysRevE.83.025102,SciRep2012_universalFeatures,1367-2630-14-1-013055}.

One of the most successful
mechanisms to model growing networks remains preferential
attachment (PA)~\cite{simon1955class,Barabasi:1999uu}. The original PA
model starts from a small seed network that grows by injecting nodes one at a
time, and each newly injected node connects to $m_0$ existing nodes. Each existing
node $i$ is chosen randomly from the current network with a probability
proportional to its degree: $\PrPA{i} = k_i / \sum_j k_j$, where $k_i$ is the
degree, or number of neighbors, of node $i$. This ``rich-get-richer'' mechanism
leads to scale-free degree distributions, $\Pr{k} \sim k^{-(1+a)}$, where the
earliest nodes will, over time, emerge as the wealthiest hubs in the network,
accruing far more links than those nodes injected at later times.  
This strong early-mover advantage is one of the most striking features of PA\@. 

PA alone cannot account for topological and statistical features observed in
real networks such as dense modular structures~\cite{Girvan11062002} and high
clustering (the abundance of triangles beyond what is expected by
chance)~\cite{watts1998collective}, and its most significant feature, the
scale-free degree distribution, collapses in equilibrium situations (in which
node injections are balanced by node removal)~\cite{PhysRevE.74.036121}.
However, the success of PA is the identification of a minimal set of mechanistic
ingredients (growth, degree-driven attachment, and thus positive feedback) that
are required to account for a universal feature abundant in many real systems.

PA has thus been the basic starting point for more complex models that
generalize the approach to include fitness
variables~\cite{bianconi2001competition,Medo:2011eu} and temporal
correlations~\cite{PhysRevLett.109.098701} to account for higher clustering and community
structure observed in real-world scale-free networks.

Here, inspired by the simplicity and generality of PA, we address
the following general question: What are the dynamic and topological consequences if 
the attachment propensity of incoming nodes is determined by
a target node's \emph{neighborhood} instead of its pure degree. Although 
this type of modification of the original PA model is small mechanistically, 
we show that the dynamic consequences are substantial. 
Our model exhibits
\emph{emergent} aging and temporally correlated dynamics, and it naturally
possesses negative feedback in the attachment propensity of existing nodes. 
Numerical investigations supported by theory show that these effects are
controlled entirely by the attachment process. No additional, artificially imposed 
rules are necessary.

We adapt the original preferential attachment network growth model in the
following way. Instead of attaching to an existing node $i$ with probability
proportional to its degree $k_i$, we attach proportional to its \emph{clustering
coefficient} (clustering attachment, or CA)
\begin{equation}
    \PrCA{i} \propto c_i^{\alpha}+ \epsilon,\quad\text{where}\quad c_i = \frac{2 \Delta_i}{k_i(k_i-1)}
\end{equation}
 is the clustering coefficient of node $i$, $\Delta_i$ is the number of links
 between neighbors of $i$ or, equivalently, the number of triangles involving
 node $i$, $\epsilon$ is a constant probability for attachment (which may be
 zero), and the exponent $\alpha$ is a parameter in our model. Other aspects of
 network growth remain the same. 
(We assume each new node attaches to $m_0=2$ existing nodes throughout; the
features are the same for $m_0>2$, but calculations become more cumbersome.)
We investigate both growing and fixed-size evolving networks. For the latter, a
random node is removed every time a new node is added.

For the original PA mechanism, the only possible ``reaction'' upon attaching to
$i$ is to increment its degree, i.e., $k_i \to k_i + 1$. For CA, however, two
reactions are possible: $(k_i \to k_i+1, \Delta_i \to \Delta_i)$ or $(k_i \to
k_i+1, \Delta_i \to \Delta_i +1)$. While the degree always grows, the number of
triangles $\Delta_i$ around $i$ depends on whether a neighbor of $i$ also
receives a new link.

These two reactions lead to the following potential changes in the clustering
coefficient of the existing node before and after the attachment:
\begin{equation}
    \delta^{(+)} c_i =  \frac{2}{k_i+1}\left(\frac{1}{k_i}-c_i\right), \qquad    \delta^{(-)} c_i =  -\frac{2}{k_i+1}c_i.
\end{equation}
Here, $\delta^{(+)} c_i$ is the change due to connecting to $i$ and a neighbor of
$i$, while $\delta^{(-)} c_i$ is the change due to connecting to $i$ and a
non-neighbor of $i$.
Even when a new triangle is formed, the clustering
coefficient after an attachment is almost always less than it was before: An
increase in $c$ after a new node's attachment is only possible if the existing
node has degree $k > 1/c$. 
This means that, in contrast to PA, the CA mechanism does not feature
rich-get-richer effects.  Instead, attaching to a node $i$ drives down $i$'s
probability for further attachments. A pure CA system will not exhibit a
power-law degree distribution because negative feedback prevents the emergence
of hub nodes. Instead, networks grown according to CA exhibit an exponential
tail in the degree distribution.
Forming new links based on the clustering coefficient provides a
particularly simple model of such negative feedback or preferential inhibition.

\begin{figure}[t!]
    \centerline{\includegraphics[]{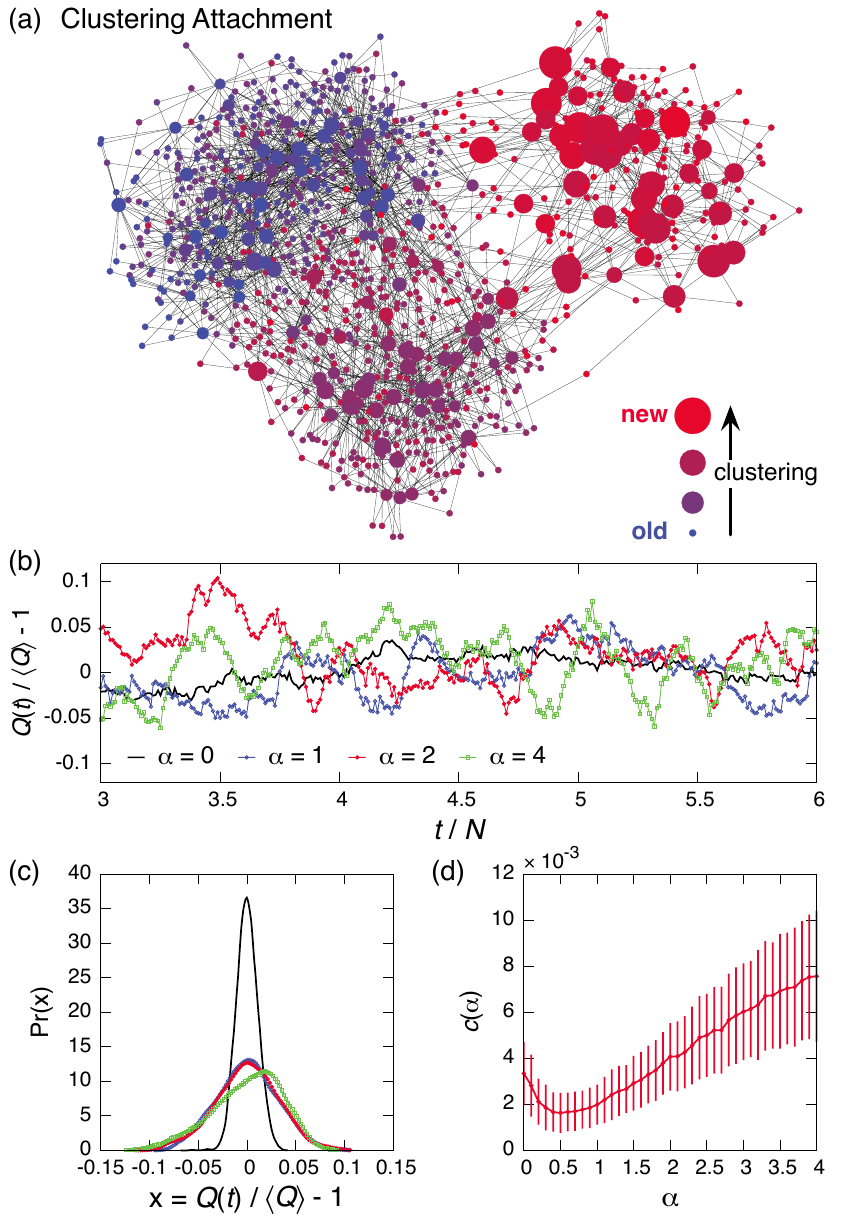}}
    \caption{(Color online) Network growth according to clustering.  %
        \lett{a}
        A realization of clustering attachment ($\alpha=2$). Node size is
        proportional to clustering and node color represents the age of the node
        (time since it was injected). Communities emerge
        approximately sequentially in time.
        \lett{b}
        A measurement of the model's community structure using modularity $Q$ by
        running a community detection algorithm~\cite{Blondel:2008vn} while a
        network evolves according to CA\@. Raw modularity scores may be
        problematic since sparsity alone can potentially force $Q$ to high
        values~\cite{Reichardt200620,PhysRevE.85.066118}. We instead consider
        $Q$ relative to $\left<Q\right>$, the average value observed over the
        course of the model. We see fluctuations in modularity over time for
        $\alpha \neq 0$ far larger than observed for purely random attachment
        ($\alpha = 0$). This quantifies the successive emergence and dissolution
        of modular structure in the model.  These fluctuations occur for both
        growing and stationary networks.
        \lett{c} 
        The relative distributions of $Q$ during the temporal evolution shown in
        (b). The random case is sharply peaked about its average value.
        \lett{d}
        The clustering coefficient averaged over all nodes, whic increases
        significantly as $\alpha$ increases.  Clustering is another hallmark of
        community structure. Error bars denote $\pm 1$s.d.
\label{fig:cartoon}
    }
\end{figure}

Yet, temporal effects play a role here as well, with the temporal sequence of
node injections determining what happens to subsequent nodes.  For example,
suppose a new node is injected and happens to form a triangle. This will give
that new node maximum $c$; it may become a \emph{hot spot} for future
attachments.
In Fig.~\ref{fig:cartoon}a we draw a single realization of the CA model with
$N=1000$ nodes and $\alpha = 2$.  Qualitatively, we observe that CA dynamics
naturally gives rise to community structure~\cite{Girvan11062002}, where the hot
spot forms the seed for a new dense group to grow. These communities tend to
form sequentially: A hot spot forms and then many nodes attach to it, driving its
attractiveness down until another seed appears.  This repeating process emerges
naturally from the attachment mechanism; nothing has been artificially imposed. 

We quantify the presence and evolution of these communities by running a community
detection method~\cite{Blondel:2008vn} as a network evolves according to CA\@.
Figure~\ref{fig:cartoon}b depicts 
the optimized modularity $Q$ of the communities
found by the method. Higher values of $Q$ can be used to indicate ``better''
communities~\cite{PhysRevE.69.026113}. However, raw values of $Q$ should be
interpreted with caution, as $Q$ can
become very large due only to sparsity in the network~\cite{Reichardt200620,PhysRevE.85.066118}.
Instead, in Fig.~\ref{fig:cartoon}b we plot modularity relative to its average
value over the evolution of each CA realization. We see distinct fluctuations in
$Q$ that are not present in the case of a random network ($\alpha=0$)
[Fig.~\ref{fig:cartoon}c].  These fluctuations are due to the sequential growth
and decay of communities: A dense community forms, boosting $Q$; then it becomes
sparser as more nodes attach to the community, lowering $Q$ until a new
community forms and the process repeats. These fluctuations are present for both
growing and stationary networks. 
Further, in Fig.~\ref{fig:cartoon}d we plot the average clustering coefficient as
a function of $\alpha$. Clustering is another hallmark of modular structure, and
it increases as $\alpha$ is increased. 
Taken together, we find that $\alpha$ plays a significant role in the modular
nature of the model.

CA thus can give rise to both correlated network structure and nontrivial
temporal dynamics. An important question, however, is if this behavior is
present for the entire range of exponents $\alpha$ or if a critical parameter
threshold exists.  To understand this and characterize the dynamics further, we
now explore (i) the aging dynamics of individual nodes after injection, and (ii)
the influence that older nodes exert on newly injected ones. 
For the latter, we fix the size of the CA networks by removing a randomly chosen
node alongside each new injection, as per Fig.~\ref{fig:cartoon}b.

\begin{figure}[t!]
    \centerline{\includegraphics[]{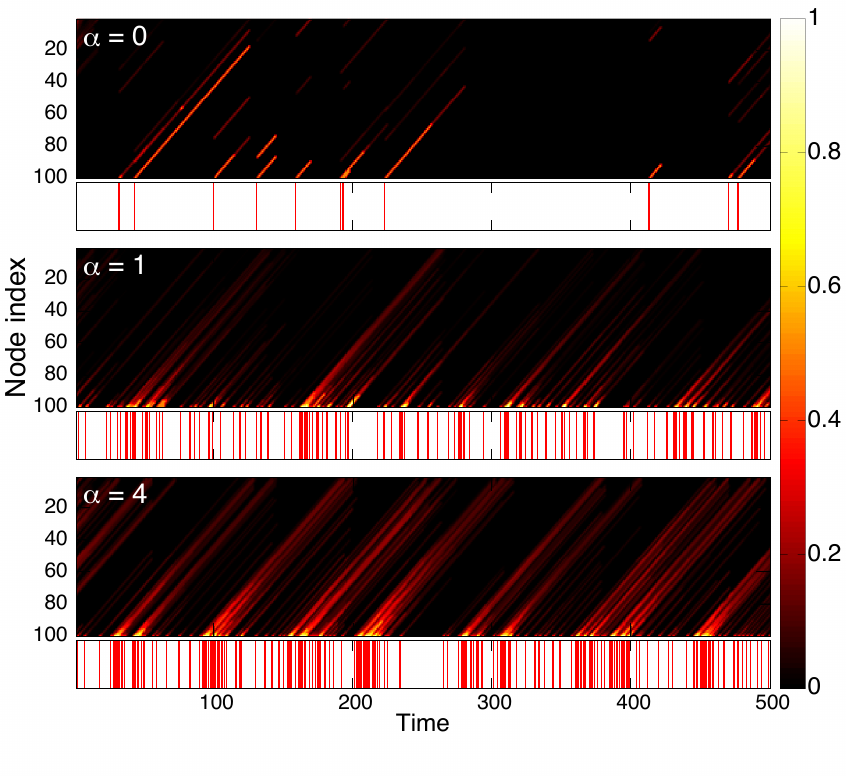}}
    \caption{(Color online) Space-time evolution for fixed-size networks of $N = 100$ nodes. 
        Each matrix element ($i,t$) represents the clustering $c_i(t)$ of node $i$ at
        time $t$.
        Nodes are indexed from oldest ($i=1$) to youngest ($i=N$). At each
        time step a new node is injected and the oldest node removed such that
        the time course of an individual node forms a diagonal across the
        matrix.
        Below each matrix is a spike train denoting injections of
        high-clustering nodes.
        As $\alpha$ increases, the clustering coefficients of individual nodes
        persist for longer times and the arrivals of high-clustering nodes
        become increasingly temporally correlated.
\label{fig:matrices}}
\end{figure}

When a new node is injected into the system, its degree $k(t)$ and clustering
$c(t)$ will evolve with the time since injection $t$.  This new node may then
exert an influence on the time course of subsequent nodes. To see this
qualitatively,  Fig.~\ref{fig:matrices} depicts ``space-time'' matrices for three
realizations of CA\@. In this matrix, each $N\times1$ column represents the
clustering coefficients of the network's nodes at that time. Nodes are ordered by age.
The oldest node is removed and a new node injected such that the time course of
$c$ for each node forms a diagonal streak across the matrix.
Below each matrix a spike train is shown, highlighting the injection times of
high-$c$ nodes.
As $\alpha$ increases, the injection times of high-$c$ nodes become temporally
correlated and the clustering coefficients of those nodes decay more slowly: 
Both temporal correlations and individual aging effects are affected
by the exponent $\alpha$ of the CA mechanism.

\begin{figure}[t!]
    \includegraphics[]{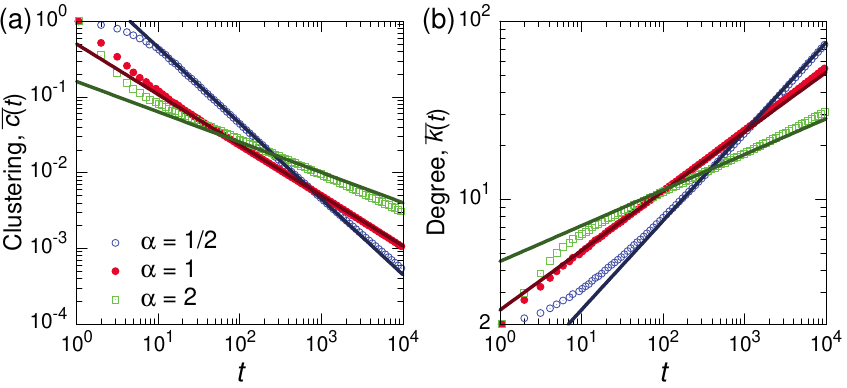}
    \caption{(Color online)  
        Expected time courses of \lett{a} clustering and \lett{b} degree as a
        function of time since injection $t$ for growing networks with different
        attachment exponents $\alpha$.  Straight lines correspond to predictions
        $\cbar(t) \sim t^{-2/(2\alpha+1)}$ and $\kbar(t) \sim
        t^{1/(2\alpha+1)}$.
        We observe the same scaling for growing and stationary systems, though
        the latter additionally feature a system-size-dependent exponential
        cutoff.
\label{fig:expectedTimeCourses}
}
\end{figure}

More quantitatively, by averaging over many realizations, we measure the
expected time courses $\cbar(t)$ and $\kbar(t)$ for nodes that are injected with
$c=1$, shown in Fig.~\ref{fig:expectedTimeCourses}. These time courses exhibit
approximate power-law decay (growth) in time for $\cbar$ ($\kbar$).

To understand the time scaling of $\cbar$ and $\kbar$, consider the following simple
analysis:
First, $\pdfrac{\kbar}{t} = \PrCAn$ and $\PrCAn \sim \cbar{(\kbar,\Delta)}^\alpha \sim
\Delta{(t)}^\alpha {\left[\kbar(t)(\kbar(t)-1)\right]}^{-\alpha} \sim \Delta^\alpha
\kbar^{-2\alpha} $.   Assuming the time evolution of $\Delta$ is approximately
constant gives $\pdfrac{\kbar}{t} \sim \kbar^{-2\alpha}$ or 
\begin{equation}
    \kbar(t) \sim t^{1/(2\alpha+1)}, \qquad \cbar(t) \sim t^{-2/(2\alpha+1)}, 
    \label{eqn:expected_k_c_scaling}
\end{equation}
where $\cbar(t)$ follows from $\cbar(t)\sim\kbar^{-2}$.
Thus we predict, if the time evolution of $\Delta$ is negligible, power-law growth in
time for degree with exponent $1 / \left(2\alpha+1\right)$ and power-law decay
in time for clustering with exponent $-2 / \left(2\alpha+1\right)$.  Despite the
simplicity of this calculation, we find good
agreement between simulations and the predicted exponents in
Eq.~\eqref{eqn:expected_k_c_scaling}; see Fig.~\ref{fig:expectedTimeCourses}.

Yet, knowing the expected temporal scaling of individual nodes' $\cbar(t)$ and
$\kbar(t)$ is insufficient to understand the emergence of the network structures
that we observe. We also need to understand the temporal nature of hot-spot
injection times. Thus, we turn to the time series of \emph{triangle injections},
or the times when nodes are introduced with $c = 1$. 
(For $m_0 > 2$, one can consider the times when new nodes appear with $c > 0$.)
These correspond to the injections of high-clustering nodes in
Fig.~\ref{fig:matrices}.

If a system displays no memory such that the probability for a spike during any time
interval ($t,t+\delta t$) depends only on $\delta t$, then the triangle
injections form a Poisson process and the interevent time, or the waiting time
between spikes, follows an exponential distribution.
Yet, many systems do not follow Poisson processes~\cite{oliveira2005human,barabasi2005origin}. 
A phenomenon is considered bursty when it possesses a memory; i.e.,\ the
probability for a new event decays with the time since the last event giving
rise to a nonexponential interevent time distribution.

\begin{figure}[t!]
    \centerline{\includegraphics[]{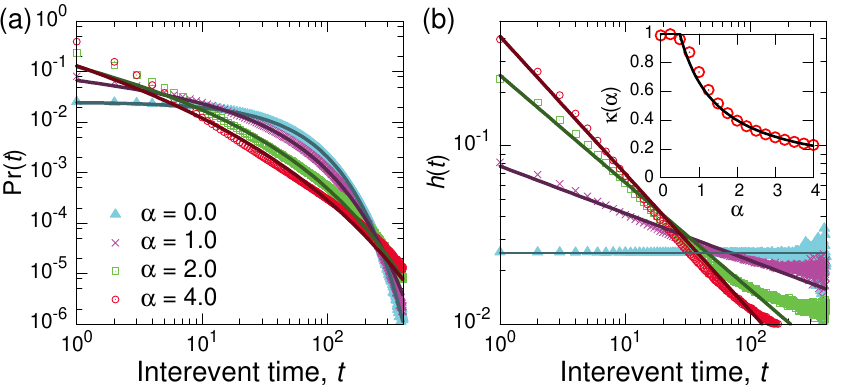}}
    \caption{(Color online) Bursty temporal features of CA\@.
        \lett{a}
        The interevent time distribution. Solid lines represent fitted Weibull
        distributions.
        \lett{b}
        The measured hazard functions and $h(t) \sim t^{\kappa-1}$.  When
        $\alpha = 0$, we recover the constant $h(t)$ ($\kappa = 1$) corresponding
        to a Poisson process.  (Inset) The observed relationship between
        $\alpha$ and the fitted $\kappa$. 
        The solid line is the prediction $\kappa = 2/(2\alpha+1)$ of
        Eq.~\eqref{eqn:summaryKappa}.
\label{fig:spike_times}}
\end{figure}

In Fig.~\ref{fig:spike_times}a, we study the interevent time distribution for
triangle injections during CA network evolution. 
(As mentioned before, to ensure the system is stationary, for the temporal
dynamics in Fig.~\ref{fig:spike_times} we now fix the size of the network by
removing one node at each time step as well.) 
When $\alpha = 0$, there is no memory and the distribution is exponential,
as expected. As $\alpha$ grows, however, the interevent time distribution becomes
more and more heavy tailed, indicating increased probability for a triangle to
form soon after a previous triangle was introduced.

A straightforward way to study bursty dynamics is through the hazard function $h(t)
= P(t) / Q(t)$, where $P(t)$ and $Q(t)$ are the probability and cumulative
distributions of waiting time $t$, respectively.
The hazard function can be interpreted as the probability rate for a new spike to
occur $t$ timesteps following the previous spike, given that no spikes occur in
the intervening time interval.
We measure the hazard functions in Fig.~\ref{fig:spike_times}b.

For a Poisson process, $h(t)$ is constant. Increasing $\alpha$ gives increasingly
non-Poissonian hazard functions: The CA mechanism naturally incorporates bursty
time dynamics in the sequences of triangle injections.

A typical property of bursty systems is a hazard function that behaves algebraically
for early times,
\begin{equation}
    h(t)  \sim t^{\kappa-1},
    \label{eqn:weibullHazard}
\end{equation}
with a singularity in continuous time for $t\rightarrow 0$. The exponent
$\kappa$ determines the degree of burstiness of the system (with $\kappa=1$
corresponding to the limiting case of a Poisson process)~\footnote{An interevent time distribution that gives
rise to such an algebraic hazard function is the Weibull distribution 
$P(t;\kappa,\lambda) = (\kappa / \lambda)  {\left(t /
\lambda\right)}^{\kappa-1}\exp\left[{-{(t/\lambda)}^\kappa}\right]$. }.

We now unify the bursty time dynamics for triangle formation with the aging time
courses for node clustering [Eq.~\eqref{eqn:expected_k_c_scaling}].
For an active system in equilibrium, the density of spikes $\rho(t)$ at time $t$
should become approximately constant (i.e.,\ independent of time) such that the
expected number of spikes emitted in a time interval $(t,t+\Delta t)$ is
proportional to $\Delta t$. (This is not the same as a Poisson process, as the
expectation is over an ensemble of CA realizations.)
Suppose a spike occurred at some past time $\tau < t$ (without loss of
generality, we shift time so that $\tau=0$).  Then, assuming spikes
are rare, a point we will return to, we approximate the spike density at $t$ by
\begin{equation}
    \rho(t) \approx \int_0^t h(s) \cbar(t-s) \, ds.
    \label{eqn:approxRho}
\end{equation}
In other words, a spike occurs at $t$, depending on the probability for the most
recent preceding spike to occur at $s$ (which is itself governed by the hazard function
for the spike at $0$) weighted by the clustering at time $t$. 

Given Eq.~\eqref{eqn:approxRho}, what hazard function will give rise to a
constant $\rho$? If $h(t) = \mathrm{const}$, we have
\begin{equation}
    \rho(t) \sim \int_0^t {(t-s)}^{-\beta} \, ds \sim t^{-\beta+1} + A,
\end{equation}
where $\beta = 2/\left(2\alpha+1\right)$ from Eq.~\eqref{eqn:expected_k_c_scaling}
and the second relation follows by introducing a constant $A$ to ensure the
initial condition $\cbar(0) = 1$ and that the integral does not diverge. 
When $\beta > 1$, $\rho(t) \to \mathrm{const}$ as $t \to \infty$, and thus we
expect an equilibrium system to be a Poisson process for $\alpha < 1/2$.

When $\beta < 1$, however, no Poisson process can be in equilibrium for our
expected $\cbar(t)$. Instead, a time-dependent hazard function $h(t) \sim
t^{\kappa-1}$ ($\kappa\neq1$) is necessary:
\begin{align}
    \rho(t) \sim \int_0^t s^{\kappa-1} {(t-s)}^{-\beta} \, ds \sim t^{\kappa-\beta},
\end{align}
where the latter holds when $\beta < 1$. 
Therefore, the system will be in equilibrium when $\kappa = \beta$.

As we mentioned, Eq.~\eqref{eqn:approxRho} is most valid at low spike densities,
where the typical time between spikes is much greater than the typical time it
takes for $\cbar(t)$ to decay.  For higher densities, the probability for a new spike to
occur at time $t$ will depend upon a superposition of earlier spikes. Yet
the contributions of the earlier spikes will \emph{each} be time independent
when $\kappa = \beta$.  Thus, our derivation should hold even at higher spike
densities.

In summary, if the above arguments hold we expect an equilibrium system to
exhibit a hazard function $h(t)
\sim t^{\kappa-1}$ with
\begin{equation}
 \kappa =
  \begin{cases}
                1                 & \text{if } \alpha < 1/2, \\
      2 / \left(2\alpha+1\right)  & \text{if } \alpha > 1/2.
  \end{cases}
  \label{eqn:summaryKappa}
\end{equation}
Indeed, there is good evidence for this relationship in the inset of
Fig.~\ref{fig:spike_times}b.

\section*{Discussion}

While positive feedback has been overwhelmingly studied in complex networks,
negative feedback remains ubiquitous in nature.
There is much room for modeling network growth besides the traditional
degree-based preferential attachment. A simple twist on this seminal work is to
form attachments based on the clustering coefficient. Doing so naturally creates
a negative feedback mechanism that leads to aging, burstiness, and the
formation of community structure in networks. The simplicity and robustness of
this mechanism is encouraging and may serve as a starting point for
investigating the origin of higher-order structures in growing networks, as well
as evolving networks that are in equilibrium. The emergence of communities and
highly variable temporal behavior observed in many complex networks, social
networks in particular, can be investigated from a CA perspective. Based on our
results, it may be promising to investigate systems in which attachment
propensities are determined by other centrality measures that capture a
different aspect of local network properties.

It is worth considering the potential practical applications of our CA model.
In a poorly understood area such as complex systems, hypothetical models such as
ours are useful for discovering potentially overlooked dynamical mechanisms and
may serve to direct future empirical studies to explore such mechanisms.
Here, one can imagine many systems where nodes are drawn not towards hubs, but
towards densely connected groups. For example, in a social network, individuals may not
want to make friends with a very popular person but, instead, with members of a
small group of very closely knit friends. 
Such hypotheses are becoming testable thanks to the appearance of
high-resolution dynamical contact networks and face-to-face proximity
data~\cite{contactNetworkPrimarySchoolPONE2011,PhysRevE.83.056109}.
Being attracted to density may also play a role in follower-followee networks for
flocking or swarming animals~\cite{nagy2010hierarchical}, where individuals may
wish to belong to a small but very cohesive group instead of being part of a
jumbled crowd all following a single leader (the hub animal).

Another area of interest may be the dynamical evolution of functional brain
networks. Indeed, positive feedback is associated with
neurological conditions such as epileptic seizures~\cite{mccormick2001cellular}. 
Recently, it has been shown~\cite{vertes10042012} that 
networks derived from fMRI data are better explained by a model where new
connections prefer to complete triangles than by traditional preferential
attachment. This model is still quite different from our work. It incorporates
anatomical distances in its attachment mechanism, but it demonstrates that
clustering can play a role in the evolution of real systems.

Preferential inhibition can also be used to model fads and fashions. For
example, music listeners may actively seek musicians that are not well known.
This corresponds to attachment probabilities that decrease with increasing
degree, of which clustering attachment is one example.

The prevalence of community structure in social systems is not explained by
degree preferential attachment alone. Likewise, social networks typically feature
exponential cutoffs in the degree distribution, simply because people have
limited time with which to maintain social relationships. This may imply that
\textbf{both} preferential attachment and preferential inhibition (or, equivalently,
density attachment) mechanisms are involved. 
Mixing some inhibition into the system will both inject community structure and
limit the formation of very high degree nodes.
Practically, this means that agents in a system are simultaneously drawn towards
highly connected regions and densely connected regions. We believe that
exploring these combined effects is a very intriguing direction for
improving our understanding of such systems.

\section*{Acknowledgments}
We thank F.~Simini and S.~Redner for many useful discussions and the
Volkswagen Foundation for support.


\begin{thebibliography}{35}%
\makeatletter
\providecommand \@ifxundefined [1]{%
 \@ifx{#1\undefined}
}%
\providecommand \@ifnum [1]{%
 \ifnum #1\expandafter \@firstoftwo
 \else \expandafter \@secondoftwo
 \fi
}%
\providecommand \@ifx [1]{%
 \ifx #1\expandafter \@firstoftwo
 \else \expandafter \@secondoftwo
 \fi
}%
\providecommand \natexlab [1]{#1}%
\providecommand \enquote  [1]{``#1''}%
\providecommand \bibnamefont  [1]{#1}%
\providecommand \bibfnamefont [1]{#1}%
\providecommand \citenamefont [1]{#1}%
\providecommand \href@noop [0]{\@secondoftwo}%
\providecommand \href [0]{\begingroup \@sanitize@url \@href}%
\providecommand \@href[1]{\@@startlink{#1}\@@href}%
\providecommand \@@href[1]{\endgroup#1\@@endlink}%
\providecommand \@sanitize@url [0]{\catcode `\\12\catcode `\$12\catcode
  `\&12\catcode `\#12\catcode `\^12\catcode `\_12\catcode `\%12\relax}%
\providecommand \@@startlink[1]{}%
\providecommand \@@endlink[0]{}%
\providecommand \url  [0]{\begingroup\@sanitize@url \@url }%
\providecommand \@url [1]{\endgroup\@href {#1}{\urlprefix }}%
\providecommand \urlprefix  [0]{URL }%
\providecommand \Eprint [0]{\href }%
\providecommand \doibase [0]{http://dx.doi.org/}%
\providecommand \selectlanguage [0]{\@gobble}%
\providecommand \bibinfo  [0]{\@secondoftwo}%
\providecommand \bibfield  [0]{\@secondoftwo}%
\providecommand \translation [1]{[#1]}%
\providecommand \BibitemOpen [0]{}%
\providecommand \bibitemStop [0]{}%
\providecommand \bibitemNoStop [0]{.\EOS\space}%
\providecommand \EOS [0]{\spacefactor3000\relax}%
\providecommand \BibitemShut  [1]{\csname bibitem#1\endcsname}%
\let\auto@bib@innerbib\@empty
\bibitem [{\citenamefont {de~Solla~Price}(1976)}]{price1976general}%
  \BibitemOpen
  \bibfield  {author} {\bibinfo {author} {\bibfnamefont {D.}~\bibnamefont
  {de~Solla~Price}},\ }\bibfield  {title} {\enquote {\bibinfo {title} {A
  general theory of bibliometric and other cumulative advantage processes},}\
  }\href@noop {} {\bibfield  {journal} {\bibinfo  {journal} {J. Am. Soc. Inf.
  Sci.}\ }\textbf {\bibinfo {volume} {27}},\ \bibinfo {pages} {292--306}
  (\bibinfo {year} {1976})}\BibitemShut {NoStop}%
\bibitem [{\citenamefont {de~Solla~Price}(1986)}]{de1986little}%
  \BibitemOpen
  \bibfield  {author} {\bibinfo {author} {\bibfnamefont {D.}~\bibnamefont
  {de~Solla~Price}},\ }\href@noop {} {\emph {\bibinfo {title} {Little science,
  big science... and beyond}}}\ (\bibinfo  {publisher} {Columbia University
  Press New York},\ \bibinfo {year} {1986})\BibitemShut {NoStop}%
\bibitem [{\citenamefont {Redner}(2005)}]{redner2005citation}%
  \BibitemOpen
  \bibfield  {author} {\bibinfo {author} {\bibfnamefont {S.}~\bibnamefont
  {Redner}},\ }\bibfield  {title} {\enquote {\bibinfo {title} {Citation
  statistics from 110 years of physical review},}\ }\href@noop {} {\bibfield
  {journal} {\bibinfo  {journal} {Physics Today}\ }\textbf {\bibinfo {volume}
  {58}},\ \bibinfo {pages} {49} (\bibinfo {year} {2005})}\BibitemShut {NoStop}%
\bibitem [{\citenamefont {Medo}\ \emph {et~al.}(2011)\citenamefont {Medo},
  \citenamefont {Cimini},\ and\ \citenamefont {Gualdi}}]{Medo:2011eu}%
  \BibitemOpen
  \bibfield  {author} {\bibinfo {author} {\bibfnamefont {M.}~\bibnamefont
  {Medo}}, \bibinfo {author} {\bibfnamefont {G.}~\bibnamefont {Cimini}}, \ and\
  \bibinfo {author} {\bibfnamefont {S.}~\bibnamefont {Gualdi}},\ }\bibfield
  {title} {\enquote {\bibinfo {title} {{Temporal effects in the growth of
  networks}},}\ }\href@noop {} {\bibfield  {journal} {\bibinfo  {journal}
  {Phys. Rev. Lett.}\ }\textbf {\bibinfo {volume} {107}},\ \bibinfo {pages}
  {238701} (\bibinfo {year} {2011})}\BibitemShut {NoStop}%
\bibitem [{\citenamefont {Eisenberg}\ and\ \citenamefont
  {Levanon}(2003)}]{eisenberg2003preferential}%
  \BibitemOpen
  \bibfield  {author} {\bibinfo {author} {\bibfnamefont {E.}~\bibnamefont
  {Eisenberg}}\ and\ \bibinfo {author} {\bibfnamefont {E.~Y.}\ \bibnamefont
  {Levanon}},\ }\bibfield  {title} {\enquote {\bibinfo {title} {Preferential
  attachment in the protein network evolution},}\ }\href@noop {} {\bibfield
  {journal} {\bibinfo  {journal} {Phys. Rev. Lett.}\ }\textbf {\bibinfo
  {volume} {91}},\ \bibinfo {pages} {138701} (\bibinfo {year}
  {2003})}\BibitemShut {NoStop}%
\bibitem [{\citenamefont {Albert}(2005)}]{albert2005scale}%
  \BibitemOpen
  \bibfield  {author} {\bibinfo {author} {\bibfnamefont {R.}~\bibnamefont
  {Albert}},\ }\bibfield  {title} {\enquote {\bibinfo {title} {Scale-free
  networks in cell biology},}\ }\href@noop {} {\bibfield  {journal} {\bibinfo
  {journal} {J. Cell. Sci.}\ }\textbf {\bibinfo {volume} {118}},\ \bibinfo
  {pages} {4947--4957} (\bibinfo {year} {2005})}\BibitemShut {NoStop}%
\bibitem [{\citenamefont {Barab{\'a}si}\ and\ \citenamefont
  {Albert}(1999)}]{Barabasi:1999uu}%
  \BibitemOpen
  \bibfield  {author} {\bibinfo {author} {\bibfnamefont {A.-L.}\ \bibnamefont
  {Barab{\'a}si}}\ and\ \bibinfo {author} {\bibfnamefont {R}~\bibnamefont
  {Albert}},\ }\bibfield  {title} {\enquote {\bibinfo {title} {{Emergence of
  scaling in random networks}},}\ }\href@noop {} {\bibfield  {journal}
  {\bibinfo  {journal} {Science}\ }\textbf {\bibinfo {volume} {286}},\ \bibinfo
  {pages} {509--512} (\bibinfo {year} {1999})}\BibitemShut {NoStop}%
\bibitem [{\citenamefont {Albert}\ and\ \citenamefont
  {Barab\'asi}(2000)}]{PhysRevLett.85.5234}%
  \BibitemOpen
  \bibfield  {author} {\bibinfo {author} {\bibfnamefont {R.}~\bibnamefont
  {Albert}}\ and\ \bibinfo {author} {\bibfnamefont {A.-L.}\ \bibnamefont
  {Barab\'asi}},\ }\bibfield  {title} {\enquote {\bibinfo {title} {Topology of
  evolving networks: Local events and universality},}\ }\href {\doibase
  10.1103/PhysRevLett.85.5234} {\bibfield  {journal} {\bibinfo  {journal}
  {Phys. Rev. Lett.}\ }\textbf {\bibinfo {volume} {85}},\ \bibinfo {pages}
  {5234--5237} (\bibinfo {year} {2000})}\BibitemShut {NoStop}%
\bibitem [{\citenamefont {Newman}(2010)}]{Newman:2010ur}%
  \BibitemOpen
  \bibfield  {author} {\bibinfo {author} {\bibfnamefont {M.~E.~J.}\
  \bibnamefont {Newman}},\ }\href@noop {} {\emph {\bibinfo {title} {{Networks:
  an introduction}}}}\ (\bibinfo  {publisher} {Oxford University Press},\
  \bibinfo {year} {2010})\BibitemShut {NoStop}%
\bibitem [{\citenamefont {Holme}\ and\ \citenamefont
  {Saram{\"a}ki}(2012)}]{Holme2012}%
  \BibitemOpen
  \bibfield  {author} {\bibinfo {author} {\bibfnamefont {P.}~\bibnamefont
  {Holme}}\ and\ \bibinfo {author} {\bibfnamefont {J.}~\bibnamefont
  {Saram{\"a}ki}},\ }\bibfield  {title} {\enquote {\bibinfo {title} {Temporal
  networks},}\ }\href {\doibase 10.1016/j.physrep.2012.03.001} {\bibfield
  {journal} {\bibinfo  {journal} {Physics Reports}\ } (\bibinfo {year}
  {2012}),\ 10.1016/j.physrep.2012.03.001}\BibitemShut {NoStop}%
\bibitem [{\citenamefont {Barabasi}(2005)}]{barabasi2005origin}%
  \BibitemOpen
  \bibfield  {author} {\bibinfo {author} {\bibfnamefont {A.-L.}\ \bibnamefont
  {Barabasi}},\ }\bibfield  {title} {\enquote {\bibinfo {title} {The origin of
  bursts and heavy tails in human dynamics},}\ }\href@noop {} {\bibfield
  {journal} {\bibinfo  {journal} {Nature}\ }\textbf {\bibinfo {volume} {435}},\
  \bibinfo {pages} {207--211} (\bibinfo {year} {2005})}\BibitemShut {NoStop}%
\bibitem [{\citenamefont {Goh}\ and\ \citenamefont
  {Barab{\'a}si}(2008)}]{goh2008burstiness}%
  \BibitemOpen
  \bibfield  {author} {\bibinfo {author} {\bibfnamefont {K.~I.}\ \bibnamefont
  {Goh}}\ and\ \bibinfo {author} {\bibfnamefont {A.-L.}\ \bibnamefont
  {Barab{\'a}si}},\ }\bibfield  {title} {\enquote {\bibinfo {title} {Burstiness
  and memory in complex systems},}\ }\href@noop {} {\bibfield  {journal}
  {\bibinfo  {journal} {EPL}\ }\textbf {\bibinfo {volume} {81}},\ \bibinfo
  {pages} {48002} (\bibinfo {year} {2008})}\BibitemShut {NoStop}%
\bibitem [{\citenamefont {Oliveira}\ and\ \citenamefont
  {Barab{\'a}si}(2005)}]{oliveira2005human}%
  \BibitemOpen
  \bibfield  {author} {\bibinfo {author} {\bibfnamefont {J.~G.}\ \bibnamefont
  {Oliveira}}\ and\ \bibinfo {author} {\bibfnamefont {A.-L.}\ \bibnamefont
  {Barab{\'a}si}},\ }\bibfield  {title} {\enquote {\bibinfo {title} {Human
  dynamics: Darwin and einstein correspondence patterns},}\ }\href@noop {}
  {\bibfield  {journal} {\bibinfo  {journal} {Nature}\ }\textbf {\bibinfo
  {volume} {437}},\ \bibinfo {pages} {1251--1251} (\bibinfo {year}
  {2005})}\BibitemShut {NoStop}%
\bibitem [{\citenamefont {Jo}\ \emph {et~al.}(2012)\citenamefont {Jo},
  \citenamefont {Karsai}, \citenamefont {Kert\'esz},\ and\ \citenamefont
  {Kaski}}]{1367-2630-14-1-013055}%
  \BibitemOpen
  \bibfield  {author} {\bibinfo {author} {\bibfnamefont {H.-H.}\ \bibnamefont
  {Jo}}, \bibinfo {author} {\bibfnamefont {M.}~\bibnamefont {Karsai}}, \bibinfo
  {author} {\bibfnamefont {J.}~\bibnamefont {Kert\'esz}}, \ and\ \bibinfo
  {author} {\bibfnamefont {K.}~\bibnamefont {Kaski}},\ }\bibfield  {title}
  {\enquote {\bibinfo {title} {Circadian pattern and burstiness in mobile phone
  communication},}\ }\href {http://stacks.iop.org/1367-2630/14/i=1/a=013055}
  {\bibfield  {journal} {\bibinfo  {journal} {New Journal of Physics}\ }\textbf
  {\bibinfo {volume} {14}},\ \bibinfo {pages} {013055} (\bibinfo {year}
  {2012})}\BibitemShut {NoStop}%
\bibitem [{\citenamefont {Granovetter}(1973)}]{granovetter1973strength}%
  \BibitemOpen
  \bibfield  {author} {\bibinfo {author} {\bibfnamefont {M.~S.}\ \bibnamefont
  {Granovetter}},\ }\bibfield  {title} {\enquote {\bibinfo {title} {The
  strength of weak ties},}\ }\href@noop {} {\bibfield  {journal} {\bibinfo
  {journal} {Amer. J. Sociol.}\ }\textbf {\bibinfo {volume} {78}},\ \bibinfo
  {pages} {1360--1380} (\bibinfo {year} {1973})}\BibitemShut {NoStop}%
\bibitem [{\citenamefont {Granovetter}(1978)}]{granovetter1978threshold}%
  \BibitemOpen
  \bibfield  {author} {\bibinfo {author} {\bibfnamefont {M.}~\bibnamefont
  {Granovetter}},\ }\bibfield  {title} {\enquote {\bibinfo {title} {Threshold
  models of collective behavior},}\ }\href@noop {} {\bibfield  {journal}
  {\bibinfo  {journal} {Amer. J. Sociol.}\ }\textbf {\bibinfo {volume} {83}},\
  \bibinfo {pages} {1420--1443} (\bibinfo {year} {1978})}\BibitemShut {NoStop}%
\bibitem [{\citenamefont {Watts}(2002)}]{watts2002simple}%
  \BibitemOpen
  \bibfield  {author} {\bibinfo {author} {\bibfnamefont {D.~J.}\ \bibnamefont
  {Watts}},\ }\bibfield  {title} {\enquote {\bibinfo {title} {A simple model of
  global cascades on random networks},}\ }\href@noop {} {\bibfield  {journal}
  {\bibinfo  {journal} {Proc. Natl. Acad. Sci. USA}\ }\textbf {\bibinfo
  {volume} {99}},\ \bibinfo {pages} {5766} (\bibinfo {year}
  {2002})}\BibitemShut {NoStop}%
\bibitem [{\citenamefont {Karsai}\ \emph {et~al.}(2011)\citenamefont {Karsai},
  \citenamefont {Kivel\"a}, \citenamefont {Pan}, \citenamefont {Kaski},
  \citenamefont {Kert\'esz}, \citenamefont {Barab\'asi},\ and\ \citenamefont
  {Saram\"aki}}]{PhysRevE.83.025102}%
  \BibitemOpen
  \bibfield  {author} {\bibinfo {author} {\bibfnamefont {M.}~\bibnamefont
  {Karsai}}, \bibinfo {author} {\bibfnamefont {M.}~\bibnamefont {Kivel\"a}},
  \bibinfo {author} {\bibfnamefont {R.~K.}\ \bibnamefont {Pan}}, \bibinfo
  {author} {\bibfnamefont {K.}~\bibnamefont {Kaski}}, \bibinfo {author}
  {\bibfnamefont {J.}~\bibnamefont {Kert\'esz}}, \bibinfo {author}
  {\bibfnamefont {A.-L.}\ \bibnamefont {Barab\'asi}}, \ and\ \bibinfo {author}
  {\bibfnamefont {J.}~\bibnamefont {Saram\"aki}},\ }\bibfield  {title}
  {\enquote {\bibinfo {title} {Small but slow world: How network topology and
  burstiness slow down spreading},}\ }\href {\doibase
  10.1103/PhysRevE.83.025102} {\bibfield  {journal} {\bibinfo  {journal} {Phys.
  Rev. E}\ }\textbf {\bibinfo {volume} {83}},\ \bibinfo {pages} {025102}
  (\bibinfo {year} {2011})}\BibitemShut {NoStop}%
\bibitem [{\citenamefont {Karsai}\ \emph {et~al.}(2012)\citenamefont {Karsai},
  \citenamefont {Kaski}, \citenamefont {Barab\'asi},\ and\ \citenamefont
  {Kert\'esz}}]{SciRep2012_universalFeatures}%
  \BibitemOpen
  \bibfield  {author} {\bibinfo {author} {\bibfnamefont {M.}~\bibnamefont
  {Karsai}}, \bibinfo {author} {\bibfnamefont {K.}~\bibnamefont {Kaski}},
  \bibinfo {author} {\bibfnamefont {A.-L.}\ \bibnamefont {Barab\'asi}}, \ and\
  \bibinfo {author} {\bibfnamefont {J.}~\bibnamefont {Kert\'esz}},\ }\bibfield
  {title} {\enquote {\bibinfo {title} {Universal features of correlated bursty
  behaviour},}\ }\href {\doibase 10.1038/srep00397} {\bibfield  {journal}
  {\bibinfo  {journal} {Sci. Rep.}\ }\textbf {\bibinfo {volume} {2}},\ \bibinfo
  {pages} {2012/05/04/online} (\bibinfo {year} {2012})}\BibitemShut {NoStop}%
\bibitem [{\citenamefont {Simon}(1955)}]{simon1955class}%
  \BibitemOpen
  \bibfield  {author} {\bibinfo {author} {\bibfnamefont {H.~A.}\ \bibnamefont
  {Simon}},\ }\bibfield  {title} {\enquote {\bibinfo {title} {On a class of
  skew distribution functions},}\ }\href@noop {} {\bibfield  {journal}
  {\bibinfo  {journal} {Biometrika}\ }\textbf {\bibinfo {volume} {42}},\
  \bibinfo {pages} {425--440} (\bibinfo {year} {1955})}\BibitemShut {NoStop}%
\bibitem [{\citenamefont {Girvan}\ and\ \citenamefont
  {Newman}(2002)}]{Girvan11062002}%
  \BibitemOpen
  \bibfield  {author} {\bibinfo {author} {\bibfnamefont {M.}~\bibnamefont
  {Girvan}}\ and\ \bibinfo {author} {\bibfnamefont {M.~E.~J.}\ \bibnamefont
  {Newman}},\ }\bibfield  {title} {\enquote {\bibinfo {title} {Community
  structure in social and biological networks},}\ }\href {\doibase
  10.1073/pnas.122653799} {\bibfield  {journal} {\bibinfo  {journal} {Proc.
  Natl. Acad. Sci. USA}\ }\textbf {\bibinfo {volume} {99}},\ \bibinfo {pages}
  {7821--7826} (\bibinfo {year} {2002})}\BibitemShut {NoStop}%
\bibitem [{\citenamefont {Watts}\ and\ \citenamefont
  {Strogatz}(1998)}]{watts1998collective}%
  \BibitemOpen
  \bibfield  {author} {\bibinfo {author} {\bibfnamefont {D.~J.}\ \bibnamefont
  {Watts}}\ and\ \bibinfo {author} {\bibfnamefont {S.~H.}\ \bibnamefont
  {Strogatz}},\ }\bibfield  {title} {\enquote {\bibinfo {title} {Collective
  dynamics of `small-world' networks},}\ }\href@noop {} {\bibfield  {journal}
  {\bibinfo  {journal} {Nature}\ }\textbf {\bibinfo {volume} {393}},\ \bibinfo
  {pages} {440--442} (\bibinfo {year} {1998})}\BibitemShut {NoStop}%
\bibitem [{\citenamefont {Moore}\ \emph {et~al.}(2006)\citenamefont {Moore},
  \citenamefont {Ghoshal},\ and\ \citenamefont {Newman}}]{PhysRevE.74.036121}%
  \BibitemOpen
  \bibfield  {author} {\bibinfo {author} {\bibfnamefont {C.}~\bibnamefont
  {Moore}}, \bibinfo {author} {\bibfnamefont {G.}~\bibnamefont {Ghoshal}}, \
  and\ \bibinfo {author} {\bibfnamefont {M.~E.~J.}\ \bibnamefont {Newman}},\
  }\bibfield  {title} {\enquote {\bibinfo {title} {Exact solutions for models
  of evolving networks with addition and deletion of nodes},}\ }\href {\doibase
  10.1103/PhysRevE.74.036121} {\bibfield  {journal} {\bibinfo  {journal} {Phys.
  Rev. E}\ }\textbf {\bibinfo {volume} {74}},\ \bibinfo {pages} {036121}
  (\bibinfo {year} {2006})}\BibitemShut {NoStop}%
\bibitem [{\citenamefont {Bianconi}\ and\ \citenamefont
  {Barab{\'a}si}(2001)}]{bianconi2001competition}%
  \BibitemOpen
  \bibfield  {author} {\bibinfo {author} {\bibfnamefont {G.}~\bibnamefont
  {Bianconi}}\ and\ \bibinfo {author} {\bibfnamefont {A.-L.}\ \bibnamefont
  {Barab{\'a}si}},\ }\bibfield  {title} {\enquote {\bibinfo {title}
  {Competition and multiscaling in evolving networks},}\ }\href@noop {}
  {\bibfield  {journal} {\bibinfo  {journal} {EPL}\ }\textbf {\bibinfo {volume}
  {54}},\ \bibinfo {pages} {436} (\bibinfo {year} {2001})}\BibitemShut
  {NoStop}%
\bibitem [{\citenamefont {Golosovsky}\ and\ \citenamefont
  {Solomon}(2012)}]{PhysRevLett.109.098701}%
  \BibitemOpen
  \bibfield  {author} {\bibinfo {author} {\bibfnamefont {M.}~\bibnamefont
  {Golosovsky}}\ and\ \bibinfo {author} {\bibfnamefont {S.}~\bibnamefont
  {Solomon}},\ }\bibfield  {title} {\enquote {\bibinfo {title} {Stochastic
  dynamical model of a growing citation network based on a self-exciting point
  process},}\ }\href {\doibase 10.1103/PhysRevLett.109.098701} {\bibfield
  {journal} {\bibinfo  {journal} {Phys. Rev. Lett.}\ }\textbf {\bibinfo
  {volume} {109}},\ \bibinfo {pages} {098701} (\bibinfo {year}
  {2012})}\BibitemShut {NoStop}%
\bibitem [{\citenamefont {Blondel}\ \emph {et~al.}(2008)\citenamefont
  {Blondel}, \citenamefont {Guillaume}, \citenamefont {Lambiotte},\ and\
  \citenamefont {Lefebvre}}]{Blondel:2008vn}%
  \BibitemOpen
  \bibfield  {author} {\bibinfo {author} {\bibfnamefont {V.~D.}\ \bibnamefont
  {Blondel}}, \bibinfo {author} {\bibfnamefont {J.~L.}\ \bibnamefont
  {Guillaume}}, \bibinfo {author} {\bibfnamefont {R}~\bibnamefont {Lambiotte}},
  \ and\ \bibinfo {author} {\bibfnamefont {E.}~\bibnamefont {Lefebvre}},\
  }\bibfield  {title} {\enquote {\bibinfo {title} {{Fast unfolding of
  communities in large networks}},}\ }\href@noop {} {\bibfield  {journal}
  {\bibinfo  {journal} {J. Stat. Mech.}\ }\textbf {\bibinfo {volume} {2008}},\
  \bibinfo {pages} {P10008} (\bibinfo {year} {2008})}\BibitemShut {NoStop}%
\bibitem [{\citenamefont {Reichardt}\ and\ \citenamefont
  {Bornholdt}(2006)}]{Reichardt200620}%
  \BibitemOpen
  \bibfield  {author} {\bibinfo {author} {\bibfnamefont {J.}~\bibnamefont
  {Reichardt}}\ and\ \bibinfo {author} {\bibfnamefont {S.}~\bibnamefont
  {Bornholdt}},\ }\bibfield  {title} {\enquote {\bibinfo {title} {When are
  networks truly modular?}}\ }\href {\doibase 10.1016/j.physd.2006.09.009}
  {\bibfield  {journal} {\bibinfo  {journal} {Physica D}\ }\textbf {\bibinfo
  {volume} {224}},\ \bibinfo {pages} {20 -- 26} (\bibinfo {year}
  {2006})}\BibitemShut {NoStop}%
\bibitem [{\citenamefont {Bagrow}(2012)}]{PhysRevE.85.066118}%
  \BibitemOpen
  \bibfield  {author} {\bibinfo {author} {\bibfnamefont {J.~P.}\ \bibnamefont
  {Bagrow}},\ }\bibfield  {title} {\enquote {\bibinfo {title} {Communities and
  bottlenecks: Trees and treelike networks have high modularity},}\ }\href
  {\doibase 10.1103/PhysRevE.85.066118} {\bibfield  {journal} {\bibinfo
  {journal} {Phys. Rev. E}\ }\textbf {\bibinfo {volume} {85}},\ \bibinfo
  {pages} {066118} (\bibinfo {year} {2012})}\BibitemShut {NoStop}%
\bibitem [{\citenamefont {Newman}\ and\ \citenamefont
  {Girvan}(2004)}]{PhysRevE.69.026113}%
  \BibitemOpen
  \bibfield  {author} {\bibinfo {author} {\bibfnamefont {M.~E.~J.}\
  \bibnamefont {Newman}}\ and\ \bibinfo {author} {\bibfnamefont
  {M.}~\bibnamefont {Girvan}},\ }\bibfield  {title} {\enquote {\bibinfo {title}
  {Finding and evaluating community structure in networks},}\ }\href {\doibase
  10.1103/PhysRevE.69.026113} {\bibfield  {journal} {\bibinfo  {journal} {Phys.
  Rev. E}\ }\textbf {\bibinfo {volume} {69}},\ \bibinfo {pages} {026113}
  (\bibinfo {year} {2004})}\BibitemShut {NoStop}%
\bibitem [{Note1()}]{Note1}%
  \BibitemOpen
  \bibinfo {note} {An interevent time distribution that gives rise to such an
  algebraic hazard function is the Weibull distribution $P(t;\kappa ,\lambda )
  = (\kappa / \lambda ) {\left (t / \lambda \right )}^{\kappa -1}\protect
  \qopname \relax o{exp}\left [{-{(t/\lambda )}^\kappa }\right ]$.}\BibitemShut
  {Stop}%
\bibitem [{\citenamefont {Stehl\'e}\ \emph {et~al.}(2011)\citenamefont
  {Stehl\'e}, \citenamefont {Voirin}, \citenamefont {Barrat}, \citenamefont
  {Cattuto}, \citenamefont {Isella}, \citenamefont {Pinton}, \citenamefont
  {Quaggiotto}, \citenamefont {Van~den Broeck}, \citenamefont {R\'egis},
  \citenamefont {Lina},\ and\ \citenamefont
  {Vanhems}}]{contactNetworkPrimarySchoolPONE2011}%
  \BibitemOpen
  \bibfield  {author} {\bibinfo {author} {\bibfnamefont {J.}~\bibnamefont
  {Stehl\'e}}, \bibinfo {author} {\bibfnamefont {N.}~\bibnamefont {Voirin}},
  \bibinfo {author} {\bibfnamefont {A.}~\bibnamefont {Barrat}}, \bibinfo
  {author} {\bibfnamefont {C.}~\bibnamefont {Cattuto}}, \bibinfo {author}
  {\bibfnamefont {L.}~\bibnamefont {Isella}}, \bibinfo {author} {\bibfnamefont
  {J.-F.}\ \bibnamefont {Pinton}}, \bibinfo {author} {\bibfnamefont
  {M.}~\bibnamefont {Quaggiotto}}, \bibinfo {author} {\bibfnamefont
  {W.}~\bibnamefont {Van~den Broeck}}, \bibinfo {author} {\bibfnamefont
  {C.}~\bibnamefont {R\'egis}}, \bibinfo {author} {\bibfnamefont
  {B.}~\bibnamefont {Lina}}, \ and\ \bibinfo {author} {\bibfnamefont
  {P.}~\bibnamefont {Vanhems}},\ }\bibfield  {title} {\enquote {\bibinfo
  {title} {High-resolution measurements of face-to-face contact patterns in a
  primary school},}\ }\href {\doibase 10.1371/journal.pone.0023176} {\bibfield
  {journal} {\bibinfo  {journal} {PLoS ONE}\ }\textbf {\bibinfo {volume} {6}},\
  \bibinfo {pages} {e23176} (\bibinfo {year} {2011})}\BibitemShut {NoStop}%
\bibitem [{\citenamefont {Zhao}\ \emph {et~al.}(2011)\citenamefont {Zhao},
  \citenamefont {Stehl\'e}, \citenamefont {Bianconi},\ and\ \citenamefont
  {Barrat}}]{PhysRevE.83.056109}%
  \BibitemOpen
  \bibfield  {author} {\bibinfo {author} {\bibfnamefont {K.}~\bibnamefont
  {Zhao}}, \bibinfo {author} {\bibfnamefont {J.}~\bibnamefont {Stehl\'e}},
  \bibinfo {author} {\bibfnamefont {G.}~\bibnamefont {Bianconi}}, \ and\
  \bibinfo {author} {\bibfnamefont {A.}~\bibnamefont {Barrat}},\ }\bibfield
  {title} {\enquote {\bibinfo {title} {Social network dynamics of face-to-face
  interactions},}\ }\href {\doibase 10.1103/PhysRevE.83.056109} {\bibfield
  {journal} {\bibinfo  {journal} {Phys. Rev. E}\ }\textbf {\bibinfo {volume}
  {83}},\ \bibinfo {pages} {056109} (\bibinfo {year} {2011})}\BibitemShut
  {NoStop}%
\bibitem [{\citenamefont {Nagy}\ \emph {et~al.}(2010)\citenamefont {Nagy},
  \citenamefont {{\'A}kos}, \citenamefont {Biro},\ and\ \citenamefont
  {Vicsek}}]{nagy2010hierarchical}%
  \BibitemOpen
  \bibfield  {author} {\bibinfo {author} {\bibfnamefont {M.}~\bibnamefont
  {Nagy}}, \bibinfo {author} {\bibfnamefont {Z.}~\bibnamefont {{\'A}kos}},
  \bibinfo {author} {\bibfnamefont {D.}~\bibnamefont {Biro}}, \ and\ \bibinfo
  {author} {\bibfnamefont {T.}~\bibnamefont {Vicsek}},\ }\bibfield  {title}
  {\enquote {\bibinfo {title} {Hierarchical group dynamics in pigeon flocks},}\
  }\href@noop {} {\bibfield  {journal} {\bibinfo  {journal} {Nature}\ }\textbf
  {\bibinfo {volume} {464}},\ \bibinfo {pages} {890--893} (\bibinfo {year}
  {2010})}\BibitemShut {NoStop}%
\bibitem [{\citenamefont {McCormick}\ and\ \citenamefont
  {Contreras}(2001)}]{mccormick2001cellular}%
  \BibitemOpen
  \bibfield  {author} {\bibinfo {author} {\bibfnamefont {D.~A.}\ \bibnamefont
  {McCormick}}\ and\ \bibinfo {author} {\bibfnamefont {D.}~\bibnamefont
  {Contreras}},\ }\bibfield  {title} {\enquote {\bibinfo {title} {On the
  cellular and network bases of epileptic seizures},}\ }\href@noop {}
  {\bibfield  {journal} {\bibinfo  {journal} {Annual Review of Physiology}\
  }\textbf {\bibinfo {volume} {63}},\ \bibinfo {pages} {815--846} (\bibinfo
  {year} {2001})}\BibitemShut {NoStop}%
\bibitem [{\citenamefont {V\'ertes}\ \emph {et~al.}(2012)\citenamefont
  {V\'ertes}, \citenamefont {Alexander-Bloch}, \citenamefont {Gogtay},
  \citenamefont {Giedd}, \citenamefont {Rapoport},\ and\ \citenamefont
  {Bullmore}}]{vertes10042012}%
  \BibitemOpen
  \bibfield  {author} {\bibinfo {author} {\bibfnamefont {P.~E.}\ \bibnamefont
  {V\'ertes}}, \bibinfo {author} {\bibfnamefont {A.~F.}\ \bibnamefont
  {Alexander-Bloch}}, \bibinfo {author} {\bibfnamefont {N.}~\bibnamefont
  {Gogtay}}, \bibinfo {author} {\bibfnamefont {J.~N.}\ \bibnamefont {Giedd}},
  \bibinfo {author} {\bibfnamefont {J.~L.}\ \bibnamefont {Rapoport}}, \ and\
  \bibinfo {author} {\bibfnamefont {E.~T.}\ \bibnamefont {Bullmore}},\
  }\bibfield  {title} {\enquote {\bibinfo {title} {Simple models of human brain
  functional networks},}\ }\href {\doibase 10.1073/pnas.1111738109} {\bibfield
  {journal} {\bibinfo  {journal} {Proc. Natl. Acad. Sci. USA}\ }\textbf
  {\bibinfo {volume} {109}},\ \bibinfo {pages} {5868--5873} (\bibinfo {year}
  {2012})}\BibitemShut {NoStop}%
\end{thebibliography}
\end{document}